\documentclass[preprint]{aastex631}

\hypersetup{linkcolor=red,citecolor=cyan,filecolor=blue,urlcolor=magenta}
\usepackage{subfigure}
\usepackage{xcolor}
\usepackage{verbatim}
\shorttitle{radio/$\gamma$-ray contemporaneous brightening of high-redshift blazar PKS 0201+113}
\shortauthors{Lei et al.}

\begin{document}

\title{A $\gamma$-Ray Emitting Blazar at Redshift 3.64: Fermi-LAT and OVRO Observations of PKS 0201+113}

\correspondingauthor{Neng-Hui Liao; Tao An; Anthony  Readhead}
\email{nhliao@gzu.edu.cn;antao@shao.ac.cn;acr@caltech.edu}

\author{Hai Lei}
\affiliation{Department of Physics and Astronomy, College of Physics, Guizhou University, Guiyang 550025, China}

\author{Ying-Kang Zhang}
\affiliation{Shanghai Astronomical Observatory, Key Laboratory of Radio Astronomy, 80 Nandan Road, 200030 Shanghai, China.}

\author{Xiong Jiang}
\affiliation{Department of Physics and Astronomy, College of Physics, Guizhou University, Guiyang 550025, China}

\author{S. Kiehlmann}
\affiliation{Institute of Astrophysics, Foundation for Research and Technology-Hellas, GR-71110 Heraklion, Greece}
\affiliation{Department of Physics, Univ. of Crete, GR-70013 Heraklion, Greece}

\author{A. C. S. Readhead}
\affiliation{Owens Valley Radio Observatory, California Institute of Technology, Pasadena, CA 91125, USA}

\author{Liang Chen}
\affiliation{Shanghai Astronomical Observatory, Key Laboratory of Radio Astronomy, 80 Nandan Road, 200030 Shanghai, China.}

\author[0000-0001-6614-3344]{Neng-Hui Liao}
\affiliation{Department of Physics and Astronomy, College of Physics, Guizhou University, Guiyang 550025, China}

\author[0000-0003-4341-0029]{Tao An}
\affiliation{Shanghai Astronomical Observatory, 80 Nandan Road, 200030 Shanghai, P. R. China}
\affiliation{Key Laboratory of Radio Astronomy and Technology, Chinese Academy of Sciences, A20 Datun Road, Beijing, 100101, P. R. China}


\begin{abstract}
High-redshift  ($z >3$) $\gamma$-ray blazars are rare, but they are crucial for our understanding of jet evolution, $\gamma$-ray production and propagation, and the growth of supermassive black holes in the early universe. A new analysis of Fermi-LAT data reveals a significant (5$\sigma$), spectrally soft ($\Gamma \simeq$ 3.0) $\gamma$-ray source in a specific 4-month epoch, cospatial with PKS 0201+113 ($z$ = 3.64). Monitoring of PKS 0201+113 at 15 GHz by the Owens Valley Radio Observatory 40 m Telescope from 2008 to 2023 shows a prominent  flare that dominates the radio light curve. The  maximum of the radio flare coincides with the $\gamma$-ray flare, strongly suggesting an association  ($\textrm{p-value}=0.023$) between the $\gamma$-ray and the radio sources. PKS 0201+113 is only the third $\gamma$-ray blazar to be identified with $z> 3.5$, and it is the first such object to be identified by the detection of quasi-simultaneous $\gamma$-ray and radio flares. The jet properties of this peculiar blazar have been investigated. A detailed study of a two-zone leptonic model  is presented that fits the broadband spectral energy distribution. An alternative scenario is also briefly discussed.  

\end{abstract}

\keywords{galaxies: active -- galaxies: high redshift -- galaxies: jets -- gamma rays: galaxies -- quasars: individual (PKS 0201+113)}

\section{Introduction} \label{sec:intro}
High-redshift blazars, with Doppler-boosted emission from  relativistic jets aligned close to the line of sight \citep{1967MNRAS.135..345R,1978Natur.276..768R,1978bllo.conf..328B,1979ApJ...231..293C,1980IAUS...92..165R,2019ARA&A..57..467B}, provide luminous beacons from early cosmic epochs \citep[e.g.,][]{2004ApJ...610L...9R,2006AJ....132.1959R,2012MNRAS.426L..91S,2014ApJ...795L..29Y,2014MNRAS.440L.111G}. Studies of high-redshift blazars are central to astrophysics, since they provide key insights into the evolution of relativistic jets, jet formation, and the dynamics of AGN activity in the early universe \citep{2010MNRAS.405..387G,2010A&ARv..18..279V}. Among thousands of extragalactic $\gamma$-ray emitters, only ten blazars beyond redshift 3 have been detected by Fermi-LAT \citep[e.g.,][]{2017ApJ...837L...5A,2018ApJ...865L..17L}, making them particularly significant. Their detected $\gamma$-ray emission not only confirms their blazar nature but also provides critical information on the jet properties. Moreover, they are valuable probes of extragalactic background light (EBL) across cosmic time \citep[e.g.,][]{2018Sci...362.1031F}.

Blazars are notable for broadband violent variability on timescales ranging from minutes to years \citep{1995ARA&A..33..163W,1997ARA&A..35..445U}. Contemporaneous $\gamma$-ray variations together with those in other windows of the electromagnetic spectrum have frequently been  detected \citep[e.g.,][]{2010Natur.463..919A,2014ApJ...783...83L,2014MNRAS.441.1899F}. Given the relatively low angular resolution of Fermi-LAT compared to radio and optical bands, identifying the low-energy counterpart of a $\gamma$-ray source can be challenging. Particularly for high-redshift blazars, which exhibit soft $\gamma$-ray spectra, and given that the spatial resolution of sub-GeV photons is lower than that of GeV photons, model localization constraints from $\gamma$-ray observations are rather loose \citep[e.g.,][]{2018ApJ...865L..17L}. Therefore, the detection of synchronized variations in multiple bands is essential in order to claim a discovery of a high-redshift $\gamma$-ray emitting blazar. However, only a few such cases have been reported \citep[e.g.,][]{2019ApJ...879L...9L,2020ApJ...898L..56L,2020ApJ...900...72L}.

PKS 0201+113 is a well studied high-redshift ($z$ = 3.64, \citealt{2023ApJS..265...25J}) blazar \citep{2009A&A...495..691M}. It exhibits high radio and hard X-ray luminosities  \citep{2008ApJS..175...97H,2020ApJ...889..164M}. Because of its bright radio emission, PKS 0201+113 has been included in several radio monitoring programs \citep{2005A&A...440..409T,2011ApJS..194...29R,2018ApJS..234...12L}. Previous attempts to detect its $\gamma$-ray emission have been unsuccessful \citep{2017ApJ...837L...5A,2020ApJ...889..164M,2020ApJ...897..177P}. Considering that $\gamma$-ray emitting blazars at high-redshifts (z $>$ 3.5) are extremely rare, in this study, we carry out a thorough investigation of the  multiwavelength variability properties of PKS 0201+113 (in Section \ref{sec:data}), using Fermi-LAT data and data in lower-energy bands, and discuss the astrophysical implications (in Section \ref{sec:discu}). Here we take a $\Lambda$CDM cosmology with $H_{0}=67~{\rm km~ s^{-1}~Mpc^{-1}}$, $\Omega_{\rm m}=0.32$, and $\Omega_{\Lambda}=0.68$ \citep{2014A&A...571A..16P}.

\section{Data Analysis and Results} \label{sec:data}
\subsection{Fermi-LAT data}
We have collected the archival Fermi-LAT data ({\tt P8R3\_SOURCE\_V3}) spanning 2008 August 4 to 2023 February 25 (i.e., MJD 54683 to 60000), covering an energy range of 100 MeV to 500 GeV. We employed {\tt Fermitools} software (version 2.0.8), along with {\tt Fermitools-data} (version 0.18) for this study. During the procedure, filtering cuts, such as zenith angle $< 90^{\circ}$ and {\tt DATA\_QUAL>0 \&\& LAT\_CONFIG==1}, were applied to the photon data by the {\tt gtselect} and {\tt gtmktime} tasks, respectively. The $\gamma$-ray flux extraction was then executed using the {\tt gtlike} task, applying an {\tt Unbinned} likelihood approach. The test statistic (TS, TS = 2$\Delta\log\zeta$, \citealt{1996ApJ...461..396M}) method was used to assess the significance of the $\gamma$-ray source, where $\zeta$ are the maximum likelihood values of different models with and without the source. Background 4FGL-DR3 sources \citep{2022ApJS..260...53A} within a 15$\degr$ radius of PKS 0201+113, together with the galactic and extragalactic $\gamma$-ray diffuse components (i.e., {\tt gll\_iem\_v07.fits} and {\tt iso\_P8R3\_SOURCE\_V3\_v1.txt}), were considered during the flux extraction. We only adjusted the parameters of inner region background sources (i.e., within a radius of $10^{\circ}$ from the target), and the normalization of the diffuse templates, while keeping the other parameters fixed at the default values. Residual TS maps were generated to check for additional background sources not in 4FGL-DR3. Because of the variability of most extragalactic $\gamma$-ray sources, and Fermi-LAT's sensitivity dependence on exposure length, weak background sources with TS $<$ 10 were excluded from the $\gamma$-ray temporal analysis.

An initial analysis of the entire dataset suggested the existence of a $\gamma$-ray source (TS = 53), not included in 4FGL-DR3, located at $0.43^{\circ}$ from the radio position of PKS 0201+113. Given the $\gamma$-ray localization radius of $0.11^{\circ}$ at 95\% confidence level (C.L.),  PKS 0201+113 is unlikely the low-energy counterpart of this particular $\gamma$-ray source.  Additionaly, several other possible $\gamma$-ray sources were also identified, but they are relatively faint (TS $\leq$ 100) and distant (more than $5^{\circ}$ away). After including these sources, we added a test $\gamma$-ray source, representative of PKS 0201+113 and modeled with a power-law function assumed as its spectral template, into the analysis source model for a subsequent likelihood analysis. The resulting TS value of $\simeq$ 1 for the test source indicates no evidence for the detectable $\gamma$-ray emission from PKS 0201+113 under these circumstances. This null result is consistent with those reported in the literature \citep{2017ApJ...837L...5A,2020ApJ...889..164M,2020ApJ...897..177P}.

\begin{figure}[ht!]
\centering
\subfigure[]{
\includegraphics[scale=0.43]{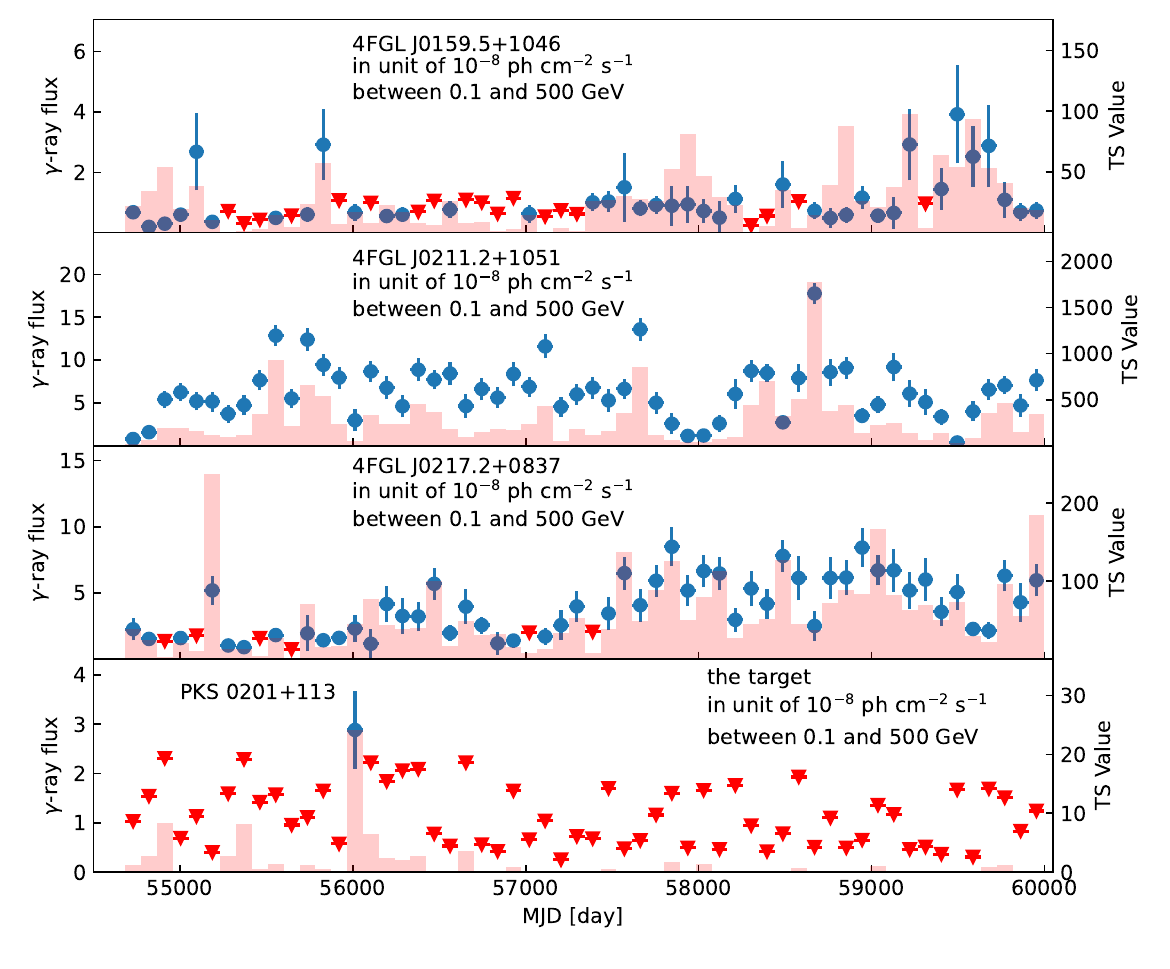}
\label{3mglc}}
\subfigure[]{
\includegraphics[scale=0.43]{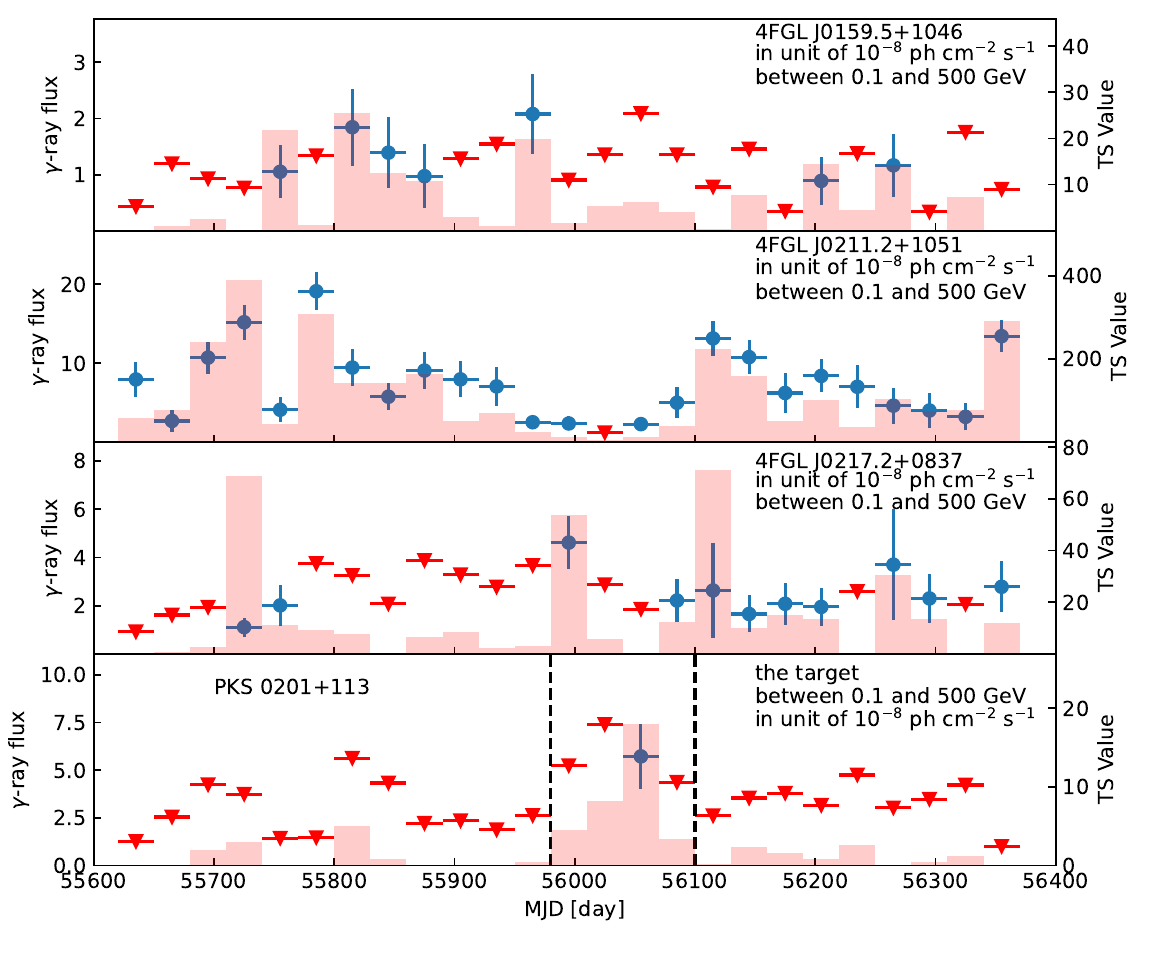}
\label{gmlc}}
\caption{$\gamma$-ray light curves of PKS 0201+113 and its neighbors. {\bf Panels (a):} 3-month time bin light curves for the entire dataset; {\bf Panels (b):} zoomed-in 1-month time bin light curves. Blue circles and red triangles are flux estimations and upper limits, respectively. Pink vertical bars are the corresponding TS values. Two dashed vertical lines are used to mark the epoch that a possible $\gamma$-ray source is seen.}

\end{figure}

We then extracted a 3-month time bin $\gamma$-ray light curve to search for any activity potentially obscured by  background emissions. Upper limits at 95\% C.L. level were calculated for low TS values ($<$ 10) using the {\tt pyLikelihood UpperLimits} tool. As shown in Figure \ref{3mglc}, one time bin is distinguished by a significant TS value ($\simeq$ 25). Since the angular resolution of Fermi-LAT for sub-GeV photons is as large as $\gtrsim~5^{\circ}$, possible contamination from nearby bright background sources has been investigated. $\gamma$-ray light curves of these sources have been extracted as well. 
The $\gamma$-ray light curves of these sources show no concurrent TS value increases, suggesting the $\gamma$-ray brightening observed in direction of PKS 0201+113 is not attributable to nearby sources. A subsequent zoomed-in 1-month time bin light curve, shown in Figure \ref{gmlc}, marks a 4-month period (i.e., MJD 55980 to 56100) of potential $\gamma$-ray activity.

\begin{figure}[ht!]
\centering
\subfigure[]{%
  \includegraphics[width=0.3\textwidth]{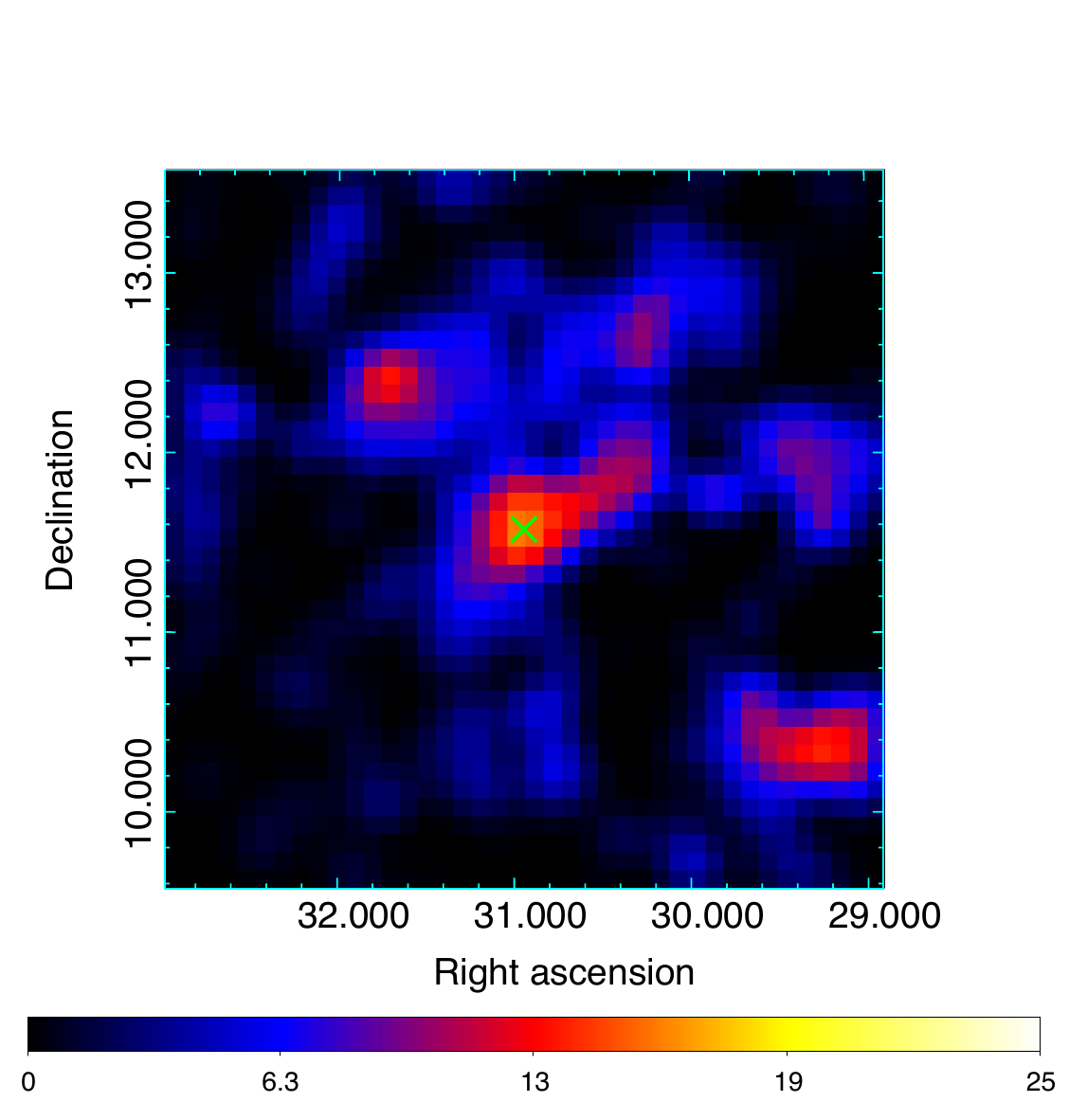}
  \label{prior}}
\subfigure[]{%
  \includegraphics[width=0.3\textwidth]{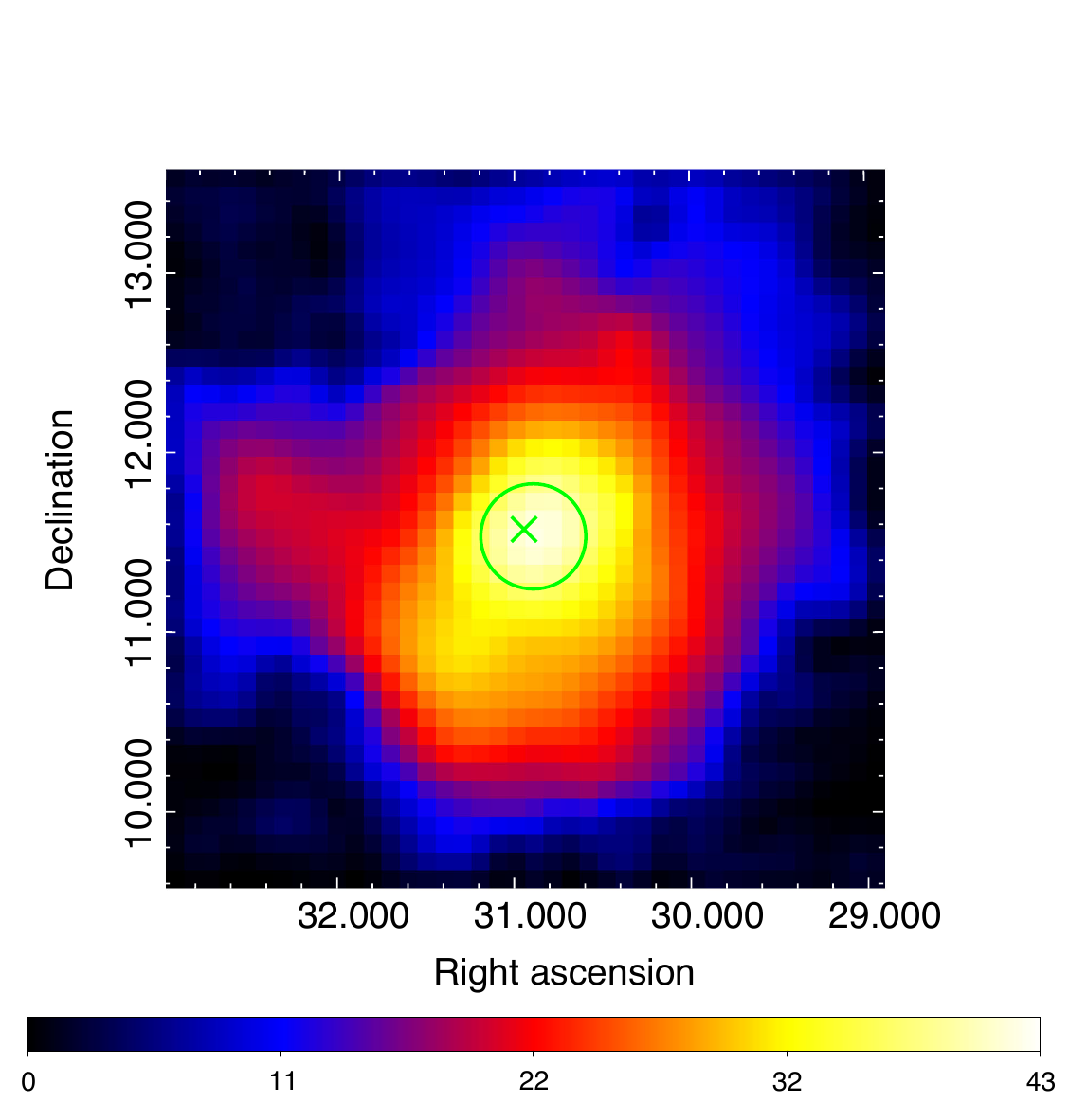}
  \label{flare}}
\subfigure[]{%
  \includegraphics[width=0.3\textwidth]{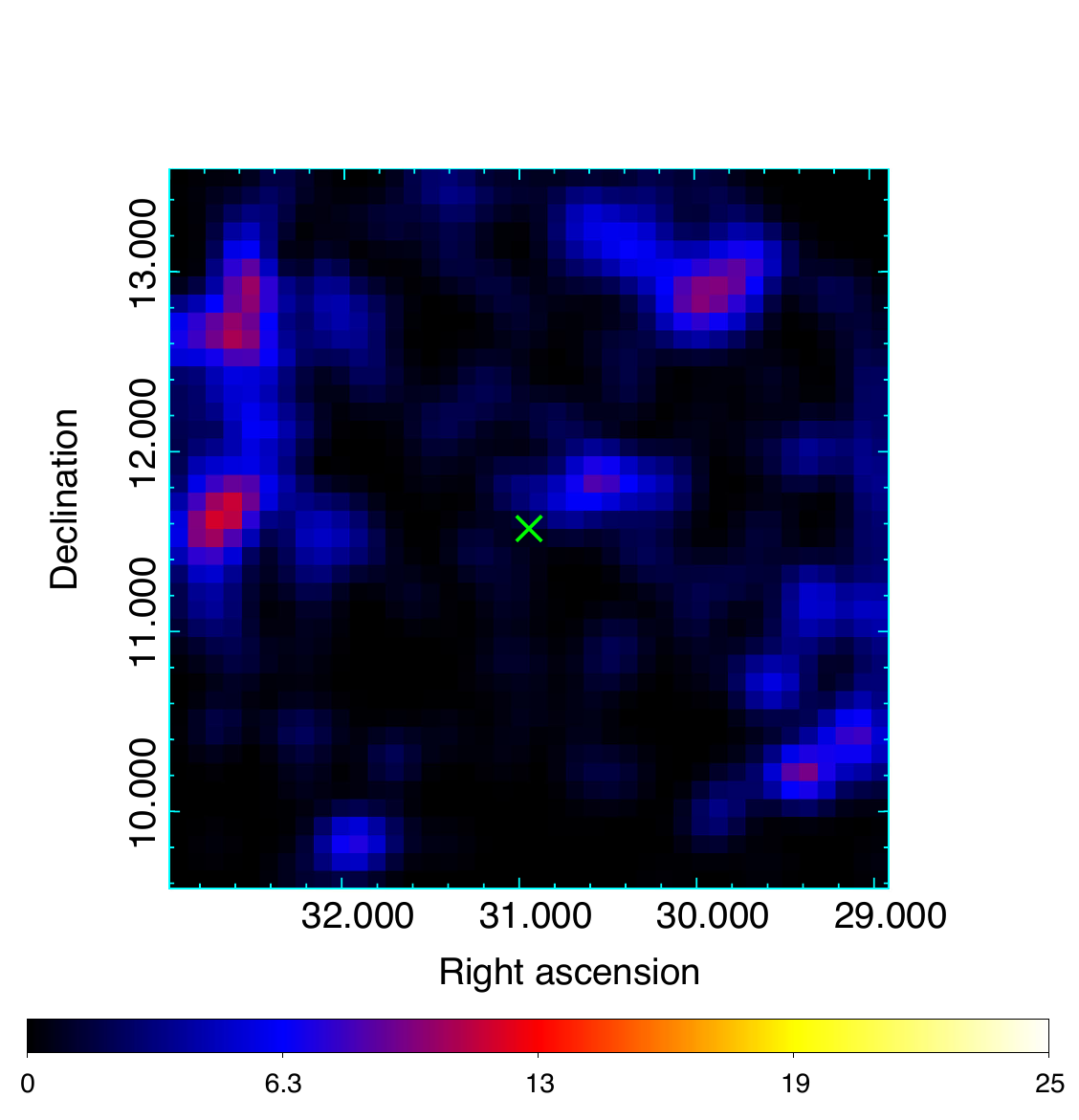}
  \label{post}}
\subfigure[]{%
  \includegraphics[scale=0.8]{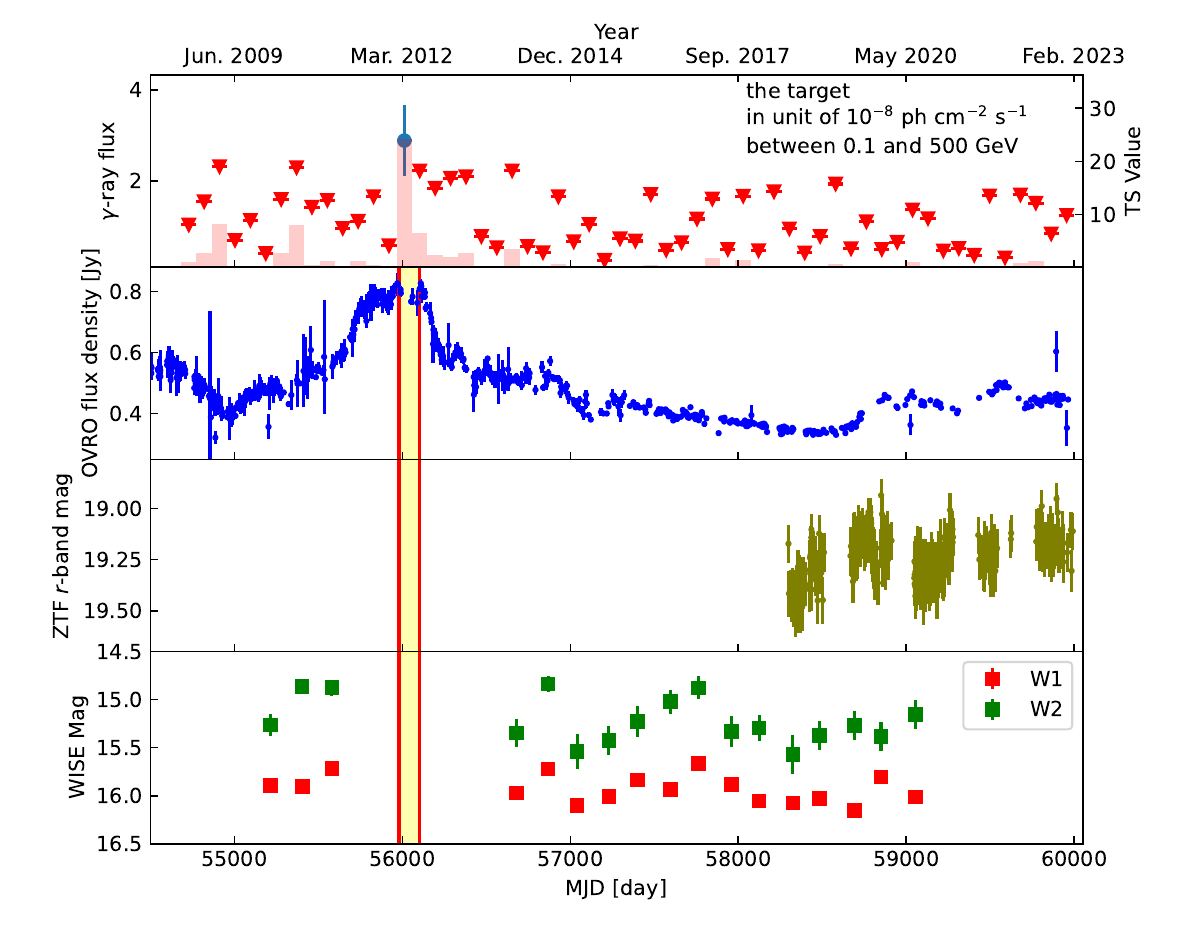}
  \label{mlc}}
\caption{
{\bf Upper panels:} smoothed $\gamma$-ray TS maps in units of degrees for different epochs (4$\degr \times 4 \degr$ scale with 0.1$\degr$ per pixel, PKS 0201+113 not included in the analysis source model). Panel (b) marks the emergence of the transient $\gamma$-ray source between MJD 55980 and 56100, in which the green circle is the 95\% C.L. error radius and the green X-shaped symbol is the radio position of PKS 0201+113. Panel (a) corresponds to the first 3.5-year Fermi-LAT data, while panel (c) corresponds to the last 10 years of data. Note that all the TS maps are centered at the optimized location of the transient $\gamma$-ray source. 
{\bf Bottom panels:} 
3-month $\gamma$-ray time bin light curve, also shown in Figure \ref{3mglc}, OVRO light curve spanning 15 years, ZTF $r$-band light curve and NEOWISE light curves. The filled yellow region marks the epoch when the transient $\gamma$-ray source appears in Figure \ref{flare}. Red vertical lines mark the start and end times of the epoch.} 
\end{figure}

Analysis focusing on this specific 4 months of data reveals a significant $\gamma$-ray source cospatial with PKS 0201+113, see Figure \ref{flare}. The optimized localization of this $\gamma$-ray source is R.A. 30.892$^{\circ}$ and DEC. 11.54$^{\circ}$, with a 95\% C. L. error radius of 0.29$^{\circ}$. The angular separation between the $\gamma$-ray location and radio position of PKS 0201+113 is 0.07$^{\circ}$. No additional blazar candidates \citep{2007ApJS..171...61H,2009A&A...495..691M}  are found within the $\gamma$-ray error radius. 

Adopting the updated $\gamma$-ray position, a single power-law function provides an acceptable description of the source. The photon flux is estimated to be $\rm (5.6 \pm 1.1) \times 10^{-8}$ ph $\rm cm^{-2}$ $\rm s^{-1}$ and the corresponding TS value is 43 (5.7$\sigma$). Considering that the 4-month epoch  was selected from a span of 14.5 years (i.e., 44 trials), the global significance after trial factor correction is 5.0$\sigma$. Our analysis shows that the $\gamma$-ray source is spectrally soft (i.e., $\Gamma$ = 2.98 $\pm$ 0.23), which supports the relatively loose localization constraints on  our model.  If this $\gamma$-ray source is indeed associated with PKS 0201+113, the corresponding $K$-corrected apparent isotropic $\gamma$-ray luminosity between 0.1 and 500~GeV is $10^{49}$ erg $\rm s^{-1}$. 

\begin{figure}[ht!]
\centering
  \includegraphics[scale=0.9]{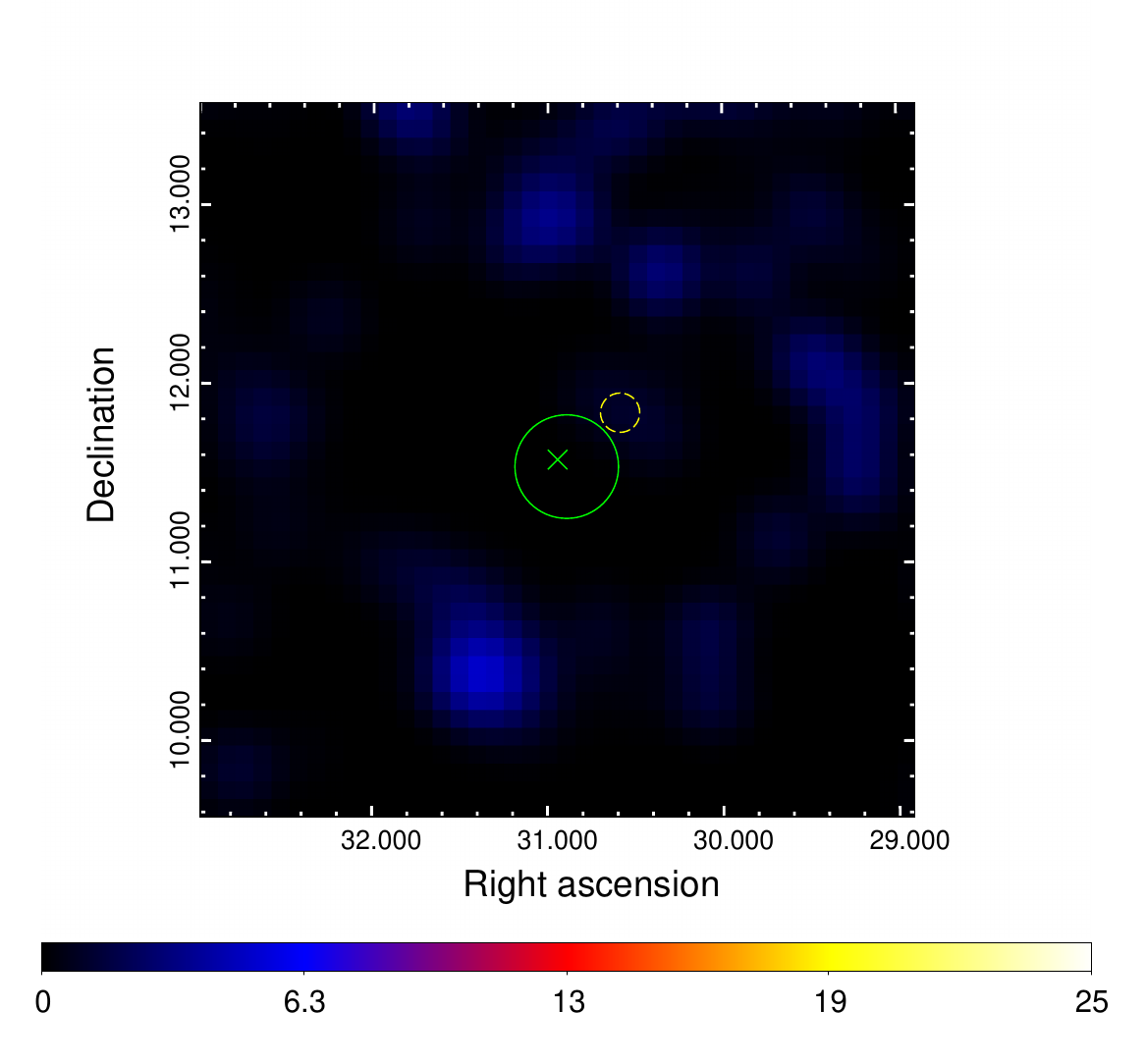}
  \label{residue}
\caption{A smoothed $\gamma$-ray TS map with the same scale and source of data as Figure \ref{flare}, but here PKS 0201+113 is included in the subtracted model to check whether there is any excess in direction of the neighbor, found in the analysis of the entire dataset. The green circle and the green X-shaped symbol are the same as in Figure \ref{flare}. The yellow circle is the 95\% C.L. error radius of the nearby source obtained from the analysis of the entire 14-yr dataset, where no significant excess is found.} 
\end{figure}

The influence  on the transient source, of the nearby neighbor from the analysis of the entire dataset has been investigated as follows. In addition to the source TS Map shown in Figure \ref{flare}, a residual TS map is shown in Figure \ref{residue}, were the contributions of the transient source and the background sources apart from the nearby neighbor have been subtracted. No significant excess (TS = 1) due to the nearby neighbor source is seen, suggesting that, at least in this specific epoch, the contribution of the nearby neighbor is not significant. 
In addition, since the source corresponding to PKS 0201+113 is highly time sensitive, we have analyzed the entire 14-yr Fermi-LAT data set, leaving out this 4-month interval,  to check whether there is a reduction of the TS of the nearby  neighbor. The analysis yields a TS value of 48, which is comparable with that derived from the intact dataset (i.e. TS = 53). This confirms that the existence  of the neighbor is irrelevant to the detection of the $\gamma$-ray source spatially coincident with PKS 0201+113.

We divided the entire dataset into three parts to investigate the long-term $\gamma$-ray variability. The analysis of the data, prior to the 4-month epoch (totally 3.5-year, see Figure \ref{prior}), suggests a tentative detection in this direction (i.e., TS = 19). The limited statistics preclude precise determination of the spectral index. The flux level during this period is moderate, $\rm (6.3 \pm 1.8) \times 10^{-9}$ ph $\rm cm^{-2}$ $\rm s^{-1}$, about one order of magnitude lower than that of the high flux state. These results are consistent with a nearby $\gamma$-ray source, 3FGL J0203.6+1148, as reported in the 3FGL catalog \citep{2015ApJS..218...23A}. However, no  low-energy counterpart for this 3FGL source is identified. Our updated localization likely benefits from the latest galactic $\gamma$-ray diffuse template. Meanwhile, analysis of the data after the flare epoch shows no sign of an excess above the background emission (TS $\leq$ 1), see Figure \ref{post}. The target remains quiescent for more than 10 years, and a 95\% C. L. upper limit of $\rm 2.1 \times 10^{-9}$ ph $\rm cm^{-2}$ $\rm s^{-1}$ is found. These findings strongly suggest that the $\gamma$-ray source exhibits substantial long-term variability. In addition, a weekly time bin light curve focused on the flare epoch was also analyzed, but  reveals no evidence of variability on timescales of several days.

\subsection{Zwicky Transient Facility Light-curve Data}
We collected the latest ZTF data \citep{2019PASP..131g8001G,2019PASP..131a8002B,2019PASP..131a8003M,ptf}\footnote{\url{https://www.ztf.caltech.edu/ztf-public-releases.html}} to analyze the target. Due to its faintness in $g$-band and sparse sampling in $i$ band, we extracted $r$-band light curve data only for objects within a 5\arcsec~radius of the target position based on the co-added reference images. Only frames with labels of {\tt catflags = 0} and {\tt chi < 4} were chosen. The snapshots were then aggregated into a  daily time bin light curve, containing 342 nights of observations from MJD 58301 to 58803, shown in Figure \ref{mlc}. No significant optical activity was found, and the standard deviation of the magnitudes ($\simeq$ 0.11 mag) is comparable to the averaged measurement uncertainty ($\simeq$ 0.10 mag).

\subsection{WISE data}
The Wide-field Infrared Survey Explorer (\emph{WISE}; \citealt{2010AJ....140.1868W}) has performed repetitive all-sky surveys in the infrared  since 2010 \citep{allwise,2014ApJ...792...30M,neowise}, except for a gap from 2011 to 2013. Time-resolved WISE/NEOWISE co-added magnitudes in the \emph{W1} and \emph{W2} bands, centered at 3.4, 4.6 $\mu$m (observed), were adopted\footnote{\url{https://github.com/fkiwy/unTimely\_Catalog\_explorer}} \citep{2014AJ....147..108L,2018AJ....156...69M,2023AJ....165...36M}. The dataset comprises 17 measurements in each infrared band between MJD 55215 and 59056, sampling every half a year except during the gap, see Figure \ref{mlc}. We applied a $\chi^{2}$-test to investigate the infrared variability, following the method of \cite{2019ApJ...879L...9L}. The analysis reveals no significant variability ($<$ 5$\sigma$) in either of the infrared light curves.

\subsection{Radio Data}
\subsubsection{Owens Valley Radio Observatory light curve}
The OVRO 40m telescope has been monitoring over 1800 objects, on a $\sim$ 3-day cadence at a frequency of 15~GHz \citep{2011ApJS..194...29R} since 2008. OVRO provides an extensive radio light curve for PKS 0201+113 spanning 15 years, see Figure \ref{mlc}. The remarkable feature in the OVRO light curve is a prominent radio flare peaking around MJD~56000. At the beginning of the OVRO observations, the radio flux density level is  $\simeq$ 0.55 Jy. This is followed by a drop to 0.40 Jy at MJD 55000, followed by a major radio flare. The flare lasts roughly 1000 days and reaches a maximum of 0.82 Jy at MJD 55974. Despite the sparse observations during the highest flux state, it is clear that the flare flux density decline starts around MJD 56109 (0.83 Jy),  and it takes roughly 185 days to drop to 0.55 Jy at MJD 56294. The flux density then decays gradually over the next $\sim$ 2300 days, reaching 0.33 Jy by MJD 58582. Subsequently, moderate activity, with peak fluxes lower than 0.5 Jy, is seen. 

The 15 GHz light curve around the time of the peak of the giant radio flare ($\gtrsim$ 0.5 Jy, between MJD 55600 and 56200) was decomposed into a series of sub-flares. The sub-flares are assumed to have an exponential shape, and they are superposed on a constant baseline flux density \citep{2002A&A...394..851S} of 0.34 Jy.  This corresponds to the observed minimum value at OVRO. As shown in Figure 5, a set of four sub-flares, together with the baseline component, fits the radio data. The peak times of the sub-flares are  MJD 55743, MJD 55833, MJD 55972 and MJD 56117, respectively, with uncertainties of several days. The rise times of the sub-flares range from a few months to nearly one year (observed), consistent with the results from studies on radio variability of Fermi blazars \citep[e.g.,][]{2011A&A...532A.146L}.

\subsubsection{VLBI radio images}
We selected the X-band Astrogeo data\footnote{\url{http://astrogeo.org}}  \citep{2009JGeod..83..859P}, focusing on epochs starting in 2011 to match the burst period around 2012. Hybrid imaging processing of the calibrated (u,v) visibility data and subsequent model fitting were performed using {\sc DIFMAP} \citep{1997ASPC..125...77S}. The corresponding parameters are presented in Table \ref{tab:imgvlbi}. Analyses of both the images and the visibility data fitting reveal that the source exhibits a typical blazar-like compact core-jet structure, similar to other high-redshift blazars \citep[e.g. J0906+6930 from ][]{2017MNRAS.468...69Z}. The VLBI data indicate the source comprises a bright core, accounting for most of the flux density, and a compact jet extending towards the northwest. Due to the insufficient (u,v) sampling from the geodetic scans and the small number of available epochs around 2012, tracking the motion of the jet components based on model-fitting results from these four epochs ws not possible. Thus only total flux densities and those of the core component are given. The uncertainties are estimated by combining the error propagation \cite[][]{1999ASPC..180..301F} with an additional 5\% of the respective flux densities.

 \subsection{Implications of $\gamma$-ray and radio variability}
A close connection between centimeter radio (i.e., observed at 8.4 and 15~GHz) and $\gamma$-ray emission of Fermi-LAT detected blazars has been recognized \citep[e.g.,][]{2009ApJ...696L..17K,2011ApJ...741...30A}. The relationship holds for mm/sub-mm radio emission \citep{2012ApJ...754...23L}. These correlations imply that the $\gamma$-ray emission likely originates from the parsec scale core-jet. Benefiting from extensive radio monitoring programs \citep[e.g.,][]{1999ApJ...512..601A,2005A&A...440..409T,2011ApJS..194...29R,2016A&A...596A..45F}, together with the wide field of view of Fermi-LAT, comparisons of temporal behaviors in the two bands have been investigated. A non-zero time delay between VLBA 15~GHz radio emission and $\gamma$-ray emission for Fermi blazars, otherwise known as $\gamma$-ray leading radio, has been detected in other objects \citep[e.g.,][]{2010ApJ...722L...7P}. The observed time lag is typically months to years, taking into account the width of the cross-correlation functions \citep[e.g.,][]{2011A&A...532A.146L,2014MNRAS.445..428M}. Interestingly, a tendency that the lag (source rest frame) decreases from cm to mm/sub-mm bands is reported, and no significant lag has been found in stacked 3mm/$\gamma$-ray emissions \citep{2014MNRAS.441.1899F}.

In PKS 0201+113, the time of emergence of the $\gamma$-ray source (i.e., MJD 55980 and 56100) coincides, to within the uncertainties, with the peak of the giant radio flare (i.e., MJD 55970 and 56109). Given that the timespan of the $\gamma$-ray and radio light curves is $>$ 5000 days, the chance probability of flares occurring  in both bands in the same 4-month interval is $\sim$ 1/44 = 0.023, which suggests a causal connection, and hence an association between the $\gamma$-ray source and the radio source.  We therefore identify PKS 0201+113  as a $\gamma$-ray emitting blazar. Due to the relatively limited statistics of the $\gamma$-ray data, cross correlation analysis of the $\gamma$-ray and OVRO light curves cannot be carried out. Nevertheless, the time bin with the largest TS value, in the monthly $\gamma$-ray light curve (Figure \ref{mlc}), is centered at MJD 56055 when the radio emission remains at a high flux level. We note that, during the period when we have optical data, the optical emission does not vary and the radio variation is also small.

The most significant observational result of this paper is the apparent near-coincidence in time between the $\gamma-$ray flare and the radio flare.  This establishes the association between these two flares, i.e. that they both originate in the same blazar.  Furthermore, it shows that the emission at these two frequencies is physically linked.  The data are insufficient to determine the exact relationship between these flares, i.e. whether the peak of the radio flare lags or precedes the $\gamma$-ray flare, although it is clear that the radio flux starts to rise around MJD 55000, well before the peak of the $\gamma-$ray flare.

The location of the jet dissipation region in blazars, where the bulk energy of the jet is converted into relativistic particles, remains an unresolved aspect of blazar physics. Rapid $\gamma$-ray variability, with timescales as short as minutes, indicates an extremely compact emitting region \citep[e.g.,][]{2007ApJ...664L..71A,2016ApJ...824L..20A}. Assuming a conical jet geometry, this region is likely to be in close proximity to the central supermassive black hole (SMBH). In that case, significant absorption of $\gamma$-ray emission via the $\gamma\gamma$ process could occur \citep{1995ApJ...441...79B,2008MNRAS.384L..19B,2010ApJ...717L.118P}. On the other hand, the location of the $\gamma-$ray emission region relative to the radio emission region can be determined via the correlated radio/$\gamma$-ray emission. The observed time lag between the observed emission in these two energy ranges is directly related to the distance between the $\gamma$-ray production site (i.e., $r_{\gamma}$) and the radio core (i.e., $r_{c}$), where the opacity of self-synchrotron absorption (SSA) is unity. The latter is inferred by the core shift measure \citep{1998A&A...330...79L}. Based on this approach, the location of the jet dissipation region could be several parsecs away from the SMBH \citep[e.g.,][]{2022MNRAS.510..469K}. 

For PKS 0201+113, unfortunately, no simultaneous VLBI images corresponding to the radio/$\gamma$-ray flares are available. Hence the size of the radio core is poorly determined. In addition, the time lag between the OVRO and $\gamma$-ray light curves is not well determined.  This makes precise localization of the jet dissipation region impossible. Nevertheless, contemporaneous brightening between the radio and $\gamma$-ray emission provides a critical clue. If the dissipation region is deeply embedded within the broad line region, a compact dissipation region leading to dense synchrotron photons and severe SSA would weaken the relationship between the emission in the $\gamma-$ray and radio bands. Conversely, if the region is significantly distant from the SMBH, beyond the edge of the dust torus, the efficiency of the external Compton (EC) process would decrease rapidly, complicating the explanation of the luminous $\gamma$-ray emission.

As mentioned above, and in spite of the papers cited above, it remains the case that the phenomenology of the correlation of $\gamma$-ray and radio variations in blazars is still largely unknown.  We know, from radio VLBI maps, that the radio emission originates in extended regions along the blazar jet, and we also know that any distance between the $\gamma$-ray emission site and the radio emission site that is deduced from time delays must be multiplied by $2 \gamma^2$ \citep{1966Natur.211..468R,1967MNRAS.135..345R} for blazars aligned within an angle $\theta < 1/\gamma$ to the line of site.  In the two cases (i.e., AO 0235+164 and B2 2308+34) where clear correlations were seen by \citet{2014MNRAS.445..428M}, we think that  an increase of activity in the blazar, caused by for example a sudden influx of new fueling \citep{1995A&A...293..665F,2023arXiv230104280G}, leads to an increase of activity and of emission which is seen first at lower energies (i.e., radio), and then, as the activity builds up it eventually leads to $\gamma$-ray emission. Thus parts of the radio flare can precede the $\gamma$-ray flare.

\section{Discussions and Summary} \label{sec:discu}
Among the thousands of sources listed in Data Release 3 of the Fourth Catalog of AGN Detected by the Fermi-LAT \citep{2022ApJS..263...24A}, only three are located beyond a redshift of 3.5, including GB 1508+5714 ($z$ = 4.3), PKS 1351-018 ($z$ = 3.72) as well as SDSS J163547.23+362930.0 ($z$ = 3.62). Additional sources identified in individual studies are PMN J2219-2719 ($z$ = 3.63, \citealt{2020ApJ...900...72L})  and NVSS J121915+365718 ($z \sim$ 3.59, \citealt{2020ApJ...903L...8P}). The identification of PKS 0201+113  ($z$ = 3.64) as a $\gamma$-ray emitter places it among the top five distant $\gamma$-ray sources known so far. Since the energy of horizon photons (i.e., $\rm \tau_{\gamma\gamma}^{EBL}$ = 1) at such a redshift is $\simeq$ 30~GeV \citep[e.g.,][]{2010ApJ...712..238F}, while the most energetic photons from PKS 0201+113 are $\sim$ 1~GeV, current EBL models are not inconsistent with our findings. The observed rather soft $\gamma$-ray spectrum of PKS 0201+113 is consistent with those of other known high-redshift $\gamma$-ray blazars \citep[e.g.,][]{2018ApJ...853..159L}. Since PKS 0201+113 is not detected by Fermi-LAT when considering the entire dataset, its flux level is obviously lower than most known high-redshift $\gamma$-ray blazars. A similar case is that of PMN J2219-2719, for which a Fermi-LAT detection is only possible when choosing a specific 5-month flaring epoch \citep{2020ApJ...900...72L}. Large $\gamma$-ray variability amplitudes, concurrent with low-energy brightening can play a crucial role in proving an association with a $\gamma$-ray source. Another noteworthy case is B3 1428+422 at z = 4.72, where a co-spatial, spectrally soft $\gamma$-ray source was reported during a 10-month epoch, yet no detections were found outside this period  \citep{2018ApJ...865L..17L}. In this case, the absence of simultaneous observations in other energy bands prevents the confirmation  of a $\gamma$-ray association.

Multiwavelength campaigns for high-redshift blazars, particularly those above a redshift of 3.5, are challenging due to the faintness of the sources when they are located at high redshifts. Beyond such a distance, so far $\gamma$-ray emissions of GB 1508+5714 and PMN J2219-2719 have been identified by contemporaneous brightenings between emissions in $\gamma$-ray and optical/near-infrared regimes \citep{2020ApJ...898L..56L,2020ApJ...900...72L}. In fact, PKS 0201+113 is the first high-redshift blazar whose $\gamma$-ray emission has been identified with the aid of a radio monitoring program, which is remarkable, given that it is one of the  brightest radio sources in the early universe ($\simeq$ 1 Jy observed at 5~GHz).  The identification of $\gamma$-ray and radio variability in PKS 0201+113 provides a unique perspective on high-redshift blazars. Radio and $\gamma$-ray emission in blazars originates almost solely from the plasma jet without significant contributions from other AGN components. However, the connection between these emission regions is complex. First: radio emission at long wavelengths tends to suffer significant SSA, which is also influenced by the distance of radiating region from the SMBH, and by the strength of the magnetic field. In addition, unlike the high energy emission, which varies rapidly, the radio emission is believed to emanate from an extended region, and it varyies over months to years \citep[e.g.,][]{2004A&A...419..485C,2014MNRAS.438.3058R,2016A&A...596A..45F}. Nevertheless, as seen in PKS 0201+113 and a few other blazars, quasi-simultaneous radio and $\gamma$-ray flares \citep{2014MNRAS.445..428M,2016A&A...586A..60K,2016ApJS..226...17L,2023ApJ...944..187W,2024MNRAS.527..882C} suggest a possible close connection between the radio and the $\gamma$-ray emission. Several other  high-redshift objects are noteworthy.  CGRaBS J0733+0456 ($z$ = 3.0,  \citealt{2019ApJ...879L...9L}), which is in the OVRO monitoring program, exhibits unprecedented quasi-simultaneous IR and $\gamma$-ray flares during the ascent phase of the radio emission. QSO J0906+6930, once proposed as the farthest blazar ($z$ = 5.48, \citealt{2004ApJ...610L...9R,2006AJ....132.1959R,2020NatCo..11..143A}), shows no significant $\gamma$-ray emission, but a tentative signal in a specific 11-day epoch has been reported \citep{2018ApJ...856..105A}, coinciding with its peak OVRO radio flux density \citep{2017MNRAS.468...69Z}. In addition, in the case of the blazar candidate PSO J047.4478+27.2992 ($z$ = 6.1), the RATAN-600 telescope has observed mild radio variability \citep{2021MNRAS.503.4662M}. These cases show the important role that the radio monitoring programs play in exploring  faint non-thermal transients in the early Universe. 

The broadband spectral energy distribution (SED) of PKS 0201+113  is presented in Figure \ref{SED}. The SED of a blazar exhibits a characteristic two-bump shape \citep{2019ARA&A..57..467B}. The first bump is attributed to synchrotron emission from relativistic electrons spiraling in a magnetic field, while the second, peaking in the $\gamma$-ray range, is generally attributed to inverse Compton (IC) emission. In the ``single zone'' model these processes involve the same electron population either scattering virtual photons, in the synchrotron case \citep{1970RvMP...42..237B}, or soft photons originating from within or from outside the jet, in  the IC case \citep[e.g.][]{1992ApJ...397L...5M,1993ApJ...416..458D,1994ApJ...421..153S,2000ApJ...545..107B}. We do not have multi-frequency radio, infrared, optical, or X-ray observations covering the flares  seen at the OVRO at 15 GHz (0.81 Jy at MJD 55990) and on the Fermi-LAT (between MJD 55980 and 56100), so it is not possible to construct a simultaneous SED.  We therefore use archival multi-frequency radio measurements taken from NED\footnote{https://ned.ipac.caltech.edu/}, and  data from NuSTAR and Swift-XRT observations at MJD 57700 \citep{2020ApJ...889..164M}, ALLWISE and SDSS magnitudes \citep{allwise,2020ApJS..249....3A} to construct the SED shown in Figure \ref{SED}. Based on these  data, the synchrotron peak frequency of PKS 0201+113  in its quiescent state likely falls in the sub-mm/infrared domain, a characteristic typical of Flat Spectrum Radio Quasars (FSRQs) \citep{2016ARA&A..54..725M}. The optical/UV emission is dominated by the accretion disk, indicative of a central SMBH with $\rm \sim$ 3$\times 10^{9}$ $\rm M_{\sun}$ \citep{2020ApJ...889..164M}. 


Since the duration of the $\gamma-$ray flare in PKS 0201+113 is $\lesssim 30$ days, whereas  the ascent phase of 15 GHz radio flare alone lasts $\sim 3$ yr, it is clear that the full extent of the radio emission regions is much more extended than that of the $\gamma-$ray emission region.  We therefore adopt a two-zone leptonic model, based on the central sub-flare of the radio light curve,  to describe the broadband SED.  In this two-zone model  an inner compact region radiating at $\gamma-$ray energies is embedded in an outer extended radio emission region, as shown in Figure \ref{cartoon}. The radius of the outer region is assumed to be   $\sim 2 \times 10^{18}$ cm, while that inner region  is assumed to be   $\sim2 \times 10^{16}$ cm. The inner region could, for example, be a mini-jet, or a narrow beam, which might correspond to a plasmoid in relativistic reconnection scenarios. This model could explain any ultra-fast $\gamma$-ray variability \citep[e.g.,][]{2009MNRAS.395L..29G,2016MNRAS.462.3325P}. Multiple mini-jets have been evoked to describe complex variability behavior \citep{2014ApJ...780...87M}.  For a review of these scenarios see \citet{2016Galax...4...37M}. A different structured jet model, the spine-jet model,  successfully explains the luminous GeV/TeV $\gamma$-ray emission of radio galaxies \citep[e.g.,][]{2005A&A...432..401G,2017ApJ...842..129C}. In these structured jet models the inner region moves faster than the outer region \citep{2016Galax...4...37M}. In our two-zone leptonic model, it is similarly assumed that the inner region has a higher bulk Lorentz factor than the outer region. In each region, the emitting electrons are assumed to be homogeneously distributed and to follow a broken power-law spectrum,
\begin{equation}
N(\gamma ) \propto \left\{ \begin{array}{ll}
                    \gamma ^{-p_1}  &  \mbox{ $\gamma_{\rm min}\leq \gamma \leq \gamma_{br}$} \\
            \gamma _{\rm br}^{p_2-p_1} \gamma ^{-p_2}  &  \mbox{ $\gamma _{\rm br}<\gamma\leq\gamma_{\rm max}$, }
           \end{array}
       \right.
\label{Ngamma}
\end{equation}
where $\rm \gamma_{br}$, $\rm \gamma _{min}$ and $\rm \gamma _{max}$ are the break, the minimum and maximum energies of the electrons, and $p_{1,2}$ are indices of the broken power-law particle distribution. 
In calculating the total radiation output of PKS 0201+113, both synchrotron and IC emission has been taken into account for both the inner and the outer emission regions. Our IC analysis includes both synchrotron self-Compton (SSC) and EC processes within each region. Additionally, IC scattering of synchrotron photons originating from one region by electrons in the other region is also taken into account. In the analysis, two important physical effects are properly incorporated: SSA and the Klein–Nishina effect in the IC scattering region. The former is crucial for understanding the radio emission, especially at lower frequencies, while the latter is significant at high energies, affecting the IC scattering cross-section. The transformations of frequency and luminosity between the jet rest frame and the observers frame are $\nu = \delta\nu^{\prime}/(1+z)$ and $\nu L_{\nu} = \delta^{4}\nu^{\prime}L^{\prime}_{\nu^{\prime}}$.  These transformations are essential for reproducing the observed spectrum, as they account for the relativistic effects due to the jet motion and the expansion of the Universe. The relative Lorentz gamma factor, ($\Gamma_{rel}$), between the inner and outer emission regions  is $\Gamma_{rel} = \Gamma_{in} \Gamma_{out}(1-\beta_{in} \beta_{out})$ \citep{2005A&A...432..401G}. 
 where the individual Lorentz gamma factors of the two regions are ($\Gamma_{in}, \Gamma_{out}$) and the corresponding velocities are  ($\beta_{in}, \beta_{out}$). This aspect of the model is vital for understanding the dynamics within the jet, and how this affects the observed emission.

The leptonic scenario gives an explanation of the high flux state SED of PKS 0201+113, assuming the parameters listed in Table \ref{inpara}.  Based on the radius and the Doppler factor of the outer emission region, the variability timescale in radio band is  $t_{obs, var} \gtrsim  (1+z)R_{out}/c\delta$, which is consistent with the OVRO light curve. Assuming a conical jet geometry, the distance between the emission region and the SMBH is approximated as $r_{\gamma} \sim \delta R_{out} \simeq$ 2 pc, in agreement with findings in the literature \citep[e.g.,][]{2014MNRAS.445..428M,2022MNRAS.510..469K}. The infrared emission from the dust torus supplies the external photon field, with an energy density of 2.5 $\times 10^{-4}$ erg $\rm cm^{-3}$ \citep{2009MNRAS.397..985G}. On this model the observed OVRO 15 GHz emission is due primarily to synchrotron radiation in the outer emission region. Based on the magnetic field strength of 0.09 Gauss, the opacity of SSA at 70 GHz in source rest frame
 (17 GHz in the jet rest frame) is found to be $\lesssim 1$, which allows for a tight connection between the radio and $\gamma-$ray
emission. Note that, at the observed frequency of 15 GHz the spectral index is $ \alpha \sim -0.6$, confirming that this is optically thin emission, and we therefore expect a tight connection between the radio and $\gamma-$ray emission.
  On the other hand, the contribution from the inner emission region to the observed OVRO emission is negligible due to severe SSA effects. The Doppler factor of 28.7, and radius of  2 $\times 10^{16}$ cm,  implies an observed variability timescale of $\simeq$ 1 day.  IC scattering by electrons of the inner emission region, of the infrared radiation from dust, is responsible for the observed $\gamma$-ray emission. Modeling of such an inner emission region   of PKS 0201+113 in a quiescent state, when no $\gamma$-ray signal is detected, has been carried out by  \citet{2020ApJ...889..164M}. The major difference between these results and our model is our significantly higher  value of the Doppler factor  and  $\rm \gamma_{br}$. This discrepancy may be attributed to the ejection of a new  emission region, leading to the injection of fresh emitting particles and consequently to $\gamma$-ray brightening. This scenario has also been proposed in multi-wavelength studies for other high-redshift blazars \citep{2019ApJ...879L...9L,2020ApJ...898L..56L}.

While the above two-zone model provides a satisfactory explanation of all the known facts about PKS 0201+113, and therefore deserves serious consideration, it is important to note that the apparent near-simultaneity of the OVRO 15 GHz peak flux density and the $\gamma-$ray, peak flux needs to be interpreted with caution. It is entirely possible that the emission at radio and $\gamma-$ray energies originates in widely-separated regions, in which case, although it is highly likely that the flares seen at radio and $\gamma-$ray energies are causally connected,   the above nested two-zone model would not apply. It should be noted that  we cannot say that the observed flares at $\gamma-$ray and radio energies are  exactly ``simultaneous'' because we are considering 30-day averages of the $\gamma-$ray flux. In the two-zone model we have broken down the large flare into sub-flares to address this issue. But the sub-flare scenario, while plausible, may not be correct.   All we can say with certainty is that the peaks in the $\gamma-$ray and radio light curves occur within the same 30-day period.  Thus the time difference between the $\gamma-$ray and the radio peaks could be as large as 30 days.   This means that the distance between the $\gamma-$ray emission region and the radio emission region could be $30/(1+3.64) \times 2 \gamma^2  c$ lt dy.   For not unreasonable values of $\gamma$ this would place the $\gamma-$ray emission site outside the radio emission site.
\vskip 1pt

A more general concern is that the assumption that the peak $\gamma-$ray emission and the peak radio emission should occur at the same time is an assumption that is not straightforward to justify.  One could have a model, for example, in which increased activity is brought about by a slow and steady increase in fueling.  On such a model, one could easily imagine that this would be seen first at low (radio) energies, so that the radio flux begins to increase first.  As more fuel is added, the activity level increases, adding steadily to the radio flux, and initiating higher energy emission.  On this model, it may be that the level of activity only becomes high enough to cause $\gamma-$ray emission some time after the increase in fueling begins.  There is no guarantee on such a model that the peaks in the $\gamma-$ray and radio emission would be simultaneous. The possibility of variations in fueling leading to enhanced radio activity \citep{1995A&A...293..665F,2023arXiv230104280G} is frequently discussed in the case of tidal disruption events, and has recently been invoked to explain the relatively short lifetimes of Type 2 compact symmetric objects (CSO-2s) \citep{2024ApJ...961..242R}. While this is a very different case to that of the  standard blazar jets discussed in this paper, it provides a clear instance in which variations in fueling the SMBH has a dramatic impact on the emission from relativistic jets.

Thus, although we have shown that the observations are consistent with the two-zone model we have adopted, the 30-day uncertainty in the time of the peak of the $\gamma-$ray emission allows the possibility that the radio and $\gamma-$ray emission regions are well separated and are not nested, as is assumed in our two-zone model.

In summary, our comprehensive multi-wavelength analysis of the high-redshift blazar PKS 0201+113 reveals several key findings.  
We identified a transient $\gamma$-ray source in a specific 4-month epoch by re-examining Fermi-LAT data. This $\gamma$-ray source, located at the radio position of PKS 0201+113 and with no other blazar candidates in the vicinity,  has a TS value of 43, corresponding to a global $5\sigma$ detection significance. The $\gamma$-ray source exhibits a spectrally soft spectrum, consistent with the behavior of other known high-redshift $\gamma$-ray emitting blazars.  Subsequent Fermi-LAT data over a decade following this epoch showed no detectable signals, which indicates  $\gamma$-ray variability exceeding a factor 20. A prominent radio flare was detected in the OVRO monitoring program in which the radio flux density reaches the maximum value approximately when the $\gamma$-ray flare is seen. The detection of quasi-contemporaneous radio/$\gamma$-ray flares strongly suggests a causal connection between the emission in these two energy bands, thus providing crucial evidence of an association between the $\gamma$-ray source and PKS 0201+113. This is the third source identified with $\gamma$-ray emission above a redshift of 3.5, and the first to be identified with the aid of a radio monitoring program, as opposed to  optical/infrared band observations. A two-zone leptonic model effectively reproduces the broadband SED of PKS 0201+113, with reasonable input parameters. Future multiwavelength campaigns, initiated by radio time-domain surveys and supplemented by simultaneous observations from infrared to $\gamma$ rays, will provide further constraints on model jet properties of this intriguing high-redshift blazar.
\clearpage

\begin{acknowledgments}
We appreciate the instructive suggestions from the anonymous referee that led to a substantial improvement of this work. Talvikki Hovatta is appreciated for the helpful comments. Lea Marcotulli is appreciated for sharing the observational SED data of PKS 0201+113. This research has made use of data obtained from the High Energy Astrophysics Science Archive Research Center (HEASARC), provided by NASA's Goddard Space Flight Center. This research has made use of the NASA/IPAC Infrared Science Archive, which is funded by the NASA and operated by the California Institute of Technology. This research has made use of data from the OVRO 40-m monitoring program (Richards, J. L. et al. 2011, ApJS, 194, 29), supported by private funding from the California Institute of Technology and the Max Planck Institute for Radio Astronomy, and by NASA grants NNX08AW31G, NNX11A043G, and NNX14AQ89G and NSF grants AST-0808050 and AST-1109911.

This work was supported in part by the National SKA Program of China (No. 2022SKA0120102, 2022SKA0130103) and NSFC under grants U2031120 and 11703093. This work was also supported in part by the Special Natural Science Fund of Guizhou University (grant No. 201911A) and the First-class Physics Promotion Programme (2019) of Guizhou University.  
SK acknowledges support from the European Research Council (ERC) under the European Unions Horizon 2020 research and innovation programme under grant agreement No.~771282.
\end{acknowledgments}

\bibliographystyle{aasjournal}
\bibliography{refs}

\begin{figure}[ht!]
\centering
  \includegraphics[scale=0.9]{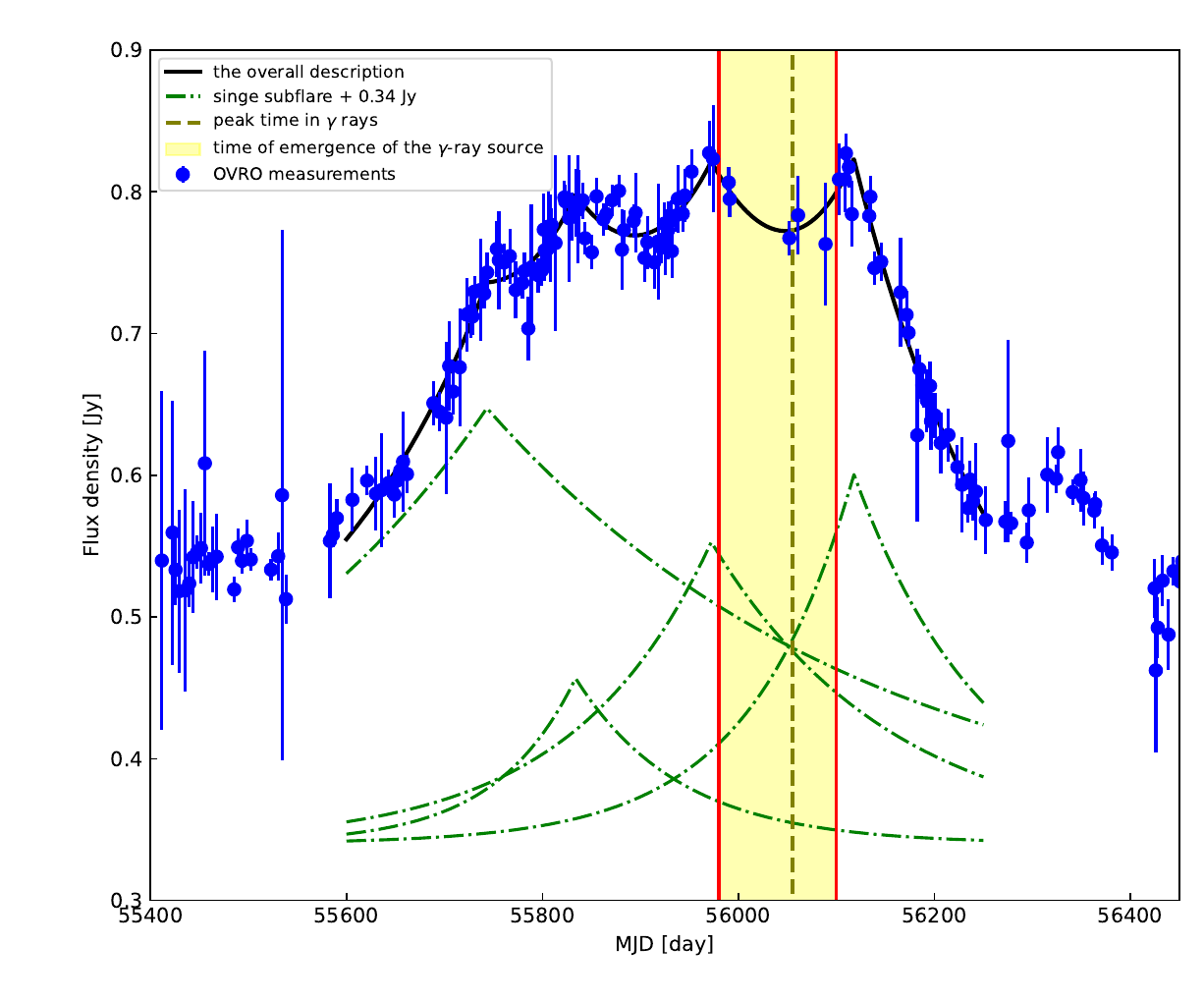}
\caption{A zoomed-in view of the OVRO light curve, focusing on the high radio flux state. The total flux density curve observed at 15 GHz has been decomposed using a series of four exponential sub-flares together with a constant baseline flux density, following the method described in \cite{2002A&A...394..851S}. The filled yellow region is as same as that shown in Figure \ref{mlc}, while the vertical dashed yellow line marks the central time of the bin with the hightest TS value in the monthly $\gamma$-ray light curve (see Figure \ref{gmlc})} 
\label{decomp}
\end{figure}

\begin{figure}
\centering
\includegraphics[scale=0.8]{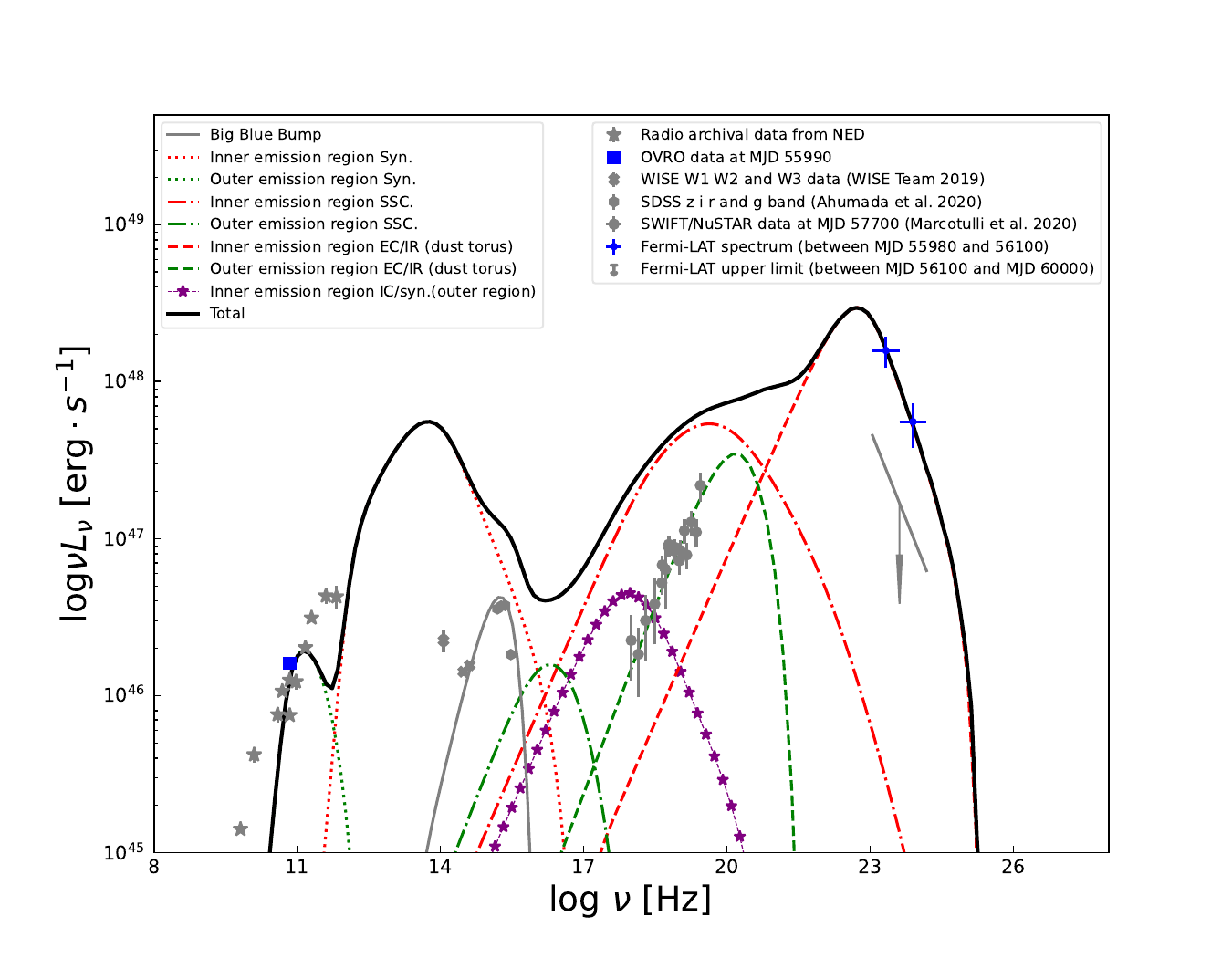}
\caption{Broadband SED of PKS 0201+113, along with the predictions of the two-zone leptonic model. Dotted, dashed dotted and dashed lines correspond to the synchrotron, SSC and EC components for the inner (colored in red) and outer emission regions (colored in green), respectively. The dashed line with purple stars is responsible for IC scattering of synchrotron emission of outer emission region by electrons of inner emission region. Radiation by interactions the other way round is too low to be seen in this diagram. The gray line represents the accretion disk emission by a central SMBH with $\rm \sim$ 3$\times 10^{9}$ $\rm M_{\sun}$ \citep{2020ApJ...889..164M}. }
\label{SED}
\end{figure}

\begin{figure}[ht!]
\centering
  \includegraphics[width=0.9\textwidth]{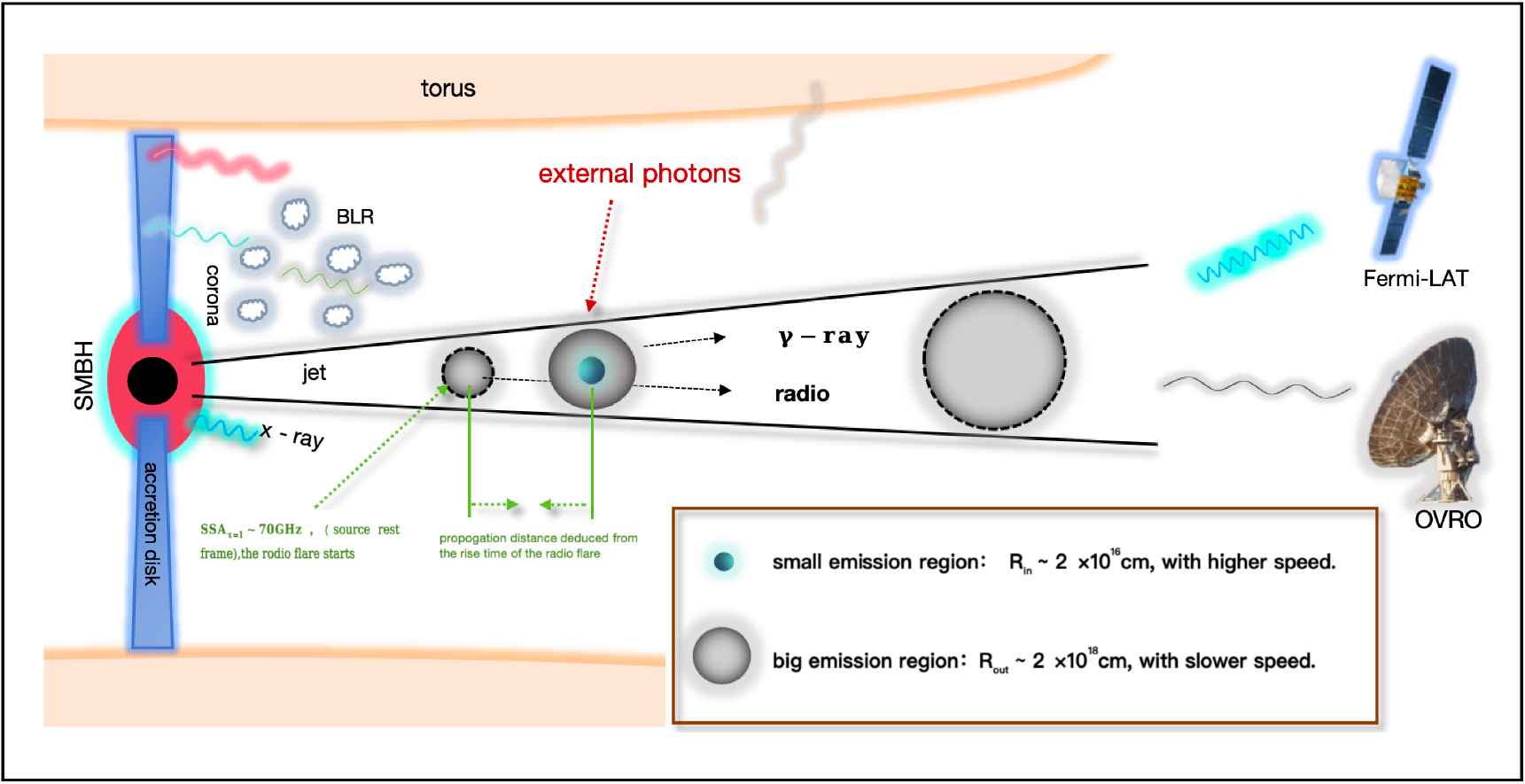}
 
\caption{A schematic illustration (not to scale) of geometry of the jet. The grey regions are responsible for generations of the radio emission, at different evolutionary stages. The left one is at the edge of the radio core where the SSA opacity is unity, and the radio flux density starts to rise. Since the emergence of the $\gamma$-ray source is at the ascent phase of a radio sub-flare peaking around MJD 56117 (see Figure \ref{decomp}), the $\gamma$-ray dissipation region (a tiny peacock green blob), embedded in an extended region (the middle grey blob), might locate downstream. The right grey region, rather distant from the central SMBH, has been plotted for an early radio activity.} 
 \label{cartoon}
\end{figure}

\clearpage
\begin{table*} 
   \centering
   \caption{VLBI information of 0203+1134}
   \begin{tabular}{ccccccccc}  \hline \hline
     Epoch  &  Freq &  $\sigma$ & S$_p$ &S$_{tot}$& S$_{core}$ & Bmaj & Bmin & Bpa  \\ 
        & (GHz) & (mJy b$^{-1}$) &  (mJy b$^{-1}$) & (mJy) & (mJy) & (mas) &(mas) & ($\degr$) \\ \hline 
       2011-05-16 & 8.4 & 1.1 & 546.7$\pm$27.4 & 741.8$\pm$37.5 & 543.0$\pm$27.2 & 2.3 & 1.0 & -4.7 \\
       2013-05-07 & 7.6 & 1.0 & 649.3$\pm$32.5 & 833.1$\pm$43.8 & 537.9$\pm$27.0 & 3.0 & 1.5 & -9.2 \\
       2014-04-01 & 8.6 & 1.0 & 489.6$\pm$24.5 & 671.4$\pm$33.9 & 409.7$\pm$20.5 & 1.7 & 0.7 & -6.9 \\
       2015-06-12 & 8.6 & 0.6 & 483.6$\pm$24.2 & 671.1$\pm$34.3 & 397.7$\pm$16.9 & 2.5 & 1.0 & -5.1 \\
\hline
   \multicolumn{9}{p{16cm}}{Notes for columns: Col.~1 -- observing date; Col.~2 -- Observing frequency; Col.~3 -- Peak flux density of the image; Col.~4 -- Total flux density of the fitted VLBI components; Col.~5 -- The flux density of the fitted VLBI core component; Col.~6-9 -- The FWHM size of the restoring beam of the image, which is also demonstrated at the bottom left corner of each image.}
   \end{tabular}
   \label{tab:imgvlbi}
\end{table*}

\begin{deluxetable}{lccccccccc}
\scriptsize
\tablenum{2} \tablewidth{0pt}
\tablecaption{List of parameters for the theoretical jet SEDs in Figure \ref{SED}}
\tablehead{ \colhead{Component} &\colhead{$p_{1}$} &\colhead{$p_{2}$} &\colhead{$\gamma_{br}$} &\colhead{$\gamma_{min}$} &\colhead{$\gamma_{max}$}  &\colhead{$B$[Gauss]} &\colhead{$\delta$} &\colhead{$R_{j}^{\prime}$[cm]}}
\startdata
Inner emission region & 1.6 & 4.5 & 700 & 1 & $\rm 10^{4}$ & 1   &  28.7 & $2 \times 10^{16}$ \\[3pt]
Outer emission region & 1.4 & 4.0 & 250 & 1 & 500 & 0.09  & 4 & $2 \times 10^{18}$ \\[3pt]
\enddata
\tablecomments{$p_{1,2}$ are the indexes of the broken power-law radiative electron distribution; $\gamma_{br}$, $\gamma_{min}$, and $\gamma_{max}$ are the break, minimum, and maximum energies of the electron distribution, respectively; B is the magnetic field strength; $\delta$ is the Doppler boosting factor; and $R^{\prime}_{j}$ is the radius of the emission emission region at the jet rest frame. \tiny} 
\label{inpara}
\end{deluxetable}

\end{document}